\documentclass[twocolumn]{article}
\usepackage{etoolbox}
\usepackage{times}

\usepackage{fancyhdr}
\usepackage{xcolor} 
\usepackage{algorithm}
\usepackage{algorithmic}
\usepackage{natbib}
\setcitestyle{numbers,square}
\usepackage{eso-pic} 
\usepackage{forloop}
\usepackage{url}

\title{Algorithmic Fairness and Color-blind Racism:  \\ Navigating the Intersection}

\author{
    Jamelle Watson-Daniels \\
    Harvard Unversity \\
    jwatsondaniels@g.harvard.edu
}

\date{}

\begin{document}

\maketitle


\begin{abstract}
Our focus lies at the intersection between two broader research perspectives: (1) the scientific study of algorithms and (2) the scholarship on race and racism. Meaningful disconnection can occur when conducting research at this intersection. What is lost in translation when algorithmic methods are applied in social contexts with ongoing problems related to systemic racism? What are opportunities for improvement when concepts from race theory are borrowed as a lens to interpret algorithmic advances? The present paper urges collective reflection on research directions at this intersection. Can we do better to engage more intentionally and thoughtfully in this space? Despite being primarily motivated by instances of racial bias, research in algorithmic fairness remains mostly disconnected from scholarship on racism. In particular, there has not been an examination connecting algorithmic fairness discussions directly to the ideology of color-blind racism; we aim to fill this gap. We see this work as a form of course correction and road building at the intersection. We begin with a review of an essential account of color-blind racism as a tool to frame this discussion. We see how ideological maneuvers can cloud and confuse how we reason about racial inequality. We review racial discourse within algorithmic fairness research and underline significant patterns, shifts and disconnects. Ultimately, we argue that researchers can improve the navigation of the landscape at the intersection by recognizing ideological shifts as such and iteratively re-orienting towards sustaining meaningful connections across interdisciplinary lines.
\end{abstract}

\maketitle

\section{Introduction}

Our focus lies at the intersection between two broader research perspectives\footnote{We define this interdisciplinary split to represent two research areas comprising multiple disciplines. We acknowledge the limitations of the dichotomy. First, there has been a good deal of overlap between the two sides in previous algorithmic fairness research papers. Second, the examples of fields listed above on each side are non-exhaustive, and there are many other axes to determine a useful split. Even so, we see this as a useful conceptualization for our discussion.}: (1) the scientific study of algorithms and (2) the scholarship on race and racism. We say that (1) includes previous work led by researchers trained in computer science, statistics, data science, etc., and (2) includes previous work led by researchers trained in race and racism, political science, investigative journalism, philosophy, etc. At the intersection, researchers primarily operating from the perspective of (1) often become interested in applying scientific methods to characterize, measure, or interpret racial disparity output by algorithms. Researchers primarily operating from the perspective of (2) often become interested in the socio-political implications of introducing algorithms into various social contexts.

Many streams of research have been born out of interest at this intersection. For instance, the realization that algorithms can perpetuate or exacerbate racial disparities in society has spurred significant research in the field of algorithmic fairness \cite{machine_bias, buolamwini2018gender, disparateinteractions, predictandserve, CaptivatingTech, fairprediction, algreparations}. More specifically, racism in algorithmic outputs has been explored in healthcare applications~\cite{panch2019artificial, cirillo2020sex} and resume filtering and hiring applications \cite{van2019hiring}. Other researchers examine racial disparities in facial recognition performance~\cite{buolamwini2018gender,10.1145/3476058,10.1145/3514094.3534153}, image captioning \cite{10.1145/3531146.3533099}, visual question answering \cite{10.1145/3531146.3533184}. In natural language processing (NLP), researchers examine discrimination towards dialects from racialized minorities such as African American English and Chicano English~\cite{ziems2022multi}, offensive speech classification \cite{10.1145/3531146.3533144, 10.1145/3531146.3533144, sap2019risk}, sentiment analysis~\cite{groenwold2020investigating,kiritchenko2018examining}, misrepresentation of dialects in speech recognition~\cite{ngueajio2022hey}, or question answering \cite{ziems2022multi,ziems2022value}, racial stereotypes in word-embeddings~\cite{manzini2019black} and text generation~\cite{sheng2019woman}.

We consider this intersection as the product of work derived from both sides. From (1) algorithms to (2) racism, the starting place might be an algorithmic question or method that is then connected to a conceptualization of racism. On the other hand, from (2) racism to (1) algorithms, the starting place might be recognizing a setting where a legacy of racism is known to persist and drawing connections between that legacy and the introduction of algorithms into the setting. As an example, \citet{buolamwini2018gender} examined disparities in classification accuracy across subgroups revealing that benchmark facial analysis algorithms had better performance for individuals identified as being male and lighter skinned. In some way, the study connected algorithmic performance metrics (error rate, true positive rate, false positive rate) to the social concepts of colorism and male privilege to highlight that the algorithm in question performed particularly poorly for  dark-skinned females. Hence, our framework would view this as an instance of work derived from (1) algorithms to (2) racism. 

Another example, the 2016 study by ProPublica of the COMPAS criminal risk assessment algorithm~\cite{machine_bias} found that Black defendants were more likely to have received a false high risk score or a false violent recidivism score, while white defendants were more likely to have recieved a false low risk score or false violent low risk score. This work~\cite{larson_angwin_kirchner_mattu_2016} raised awareness of and interest in the intersections between racism and algorithms. These types of studies focused on racial bias in risk assessments have mainly been the result of anticipating the impact of introducing algorithms into the US criminal justice system. In the US carceral system, researchers have well-documented the racist origins of mass incarceration fueled by the fact that black and brown people are over-criminalized, over-policed, and incarcerated at inhumane rates~\cite{AlexanderMichelle2010TnJC}. Given this context, the study of racial disparities with respect to risk assessment algorithms can be understood in our framework as translating from (2) racism to (1) algorithms. 

In either direction, meaningful disconnection can occur when conducting research at the intersection of racism and algorithms. What is lost in translation when algorithmic methods are applied in social contexts with ongoing problems related to systemic racism? What are opportunities for improvement when concepts from race theory are borrowed as a lens to interpret algorithmic advances? For algorithmic fairness in particular, essential disconnects noted in prior work are the disconnect from socio-political and historical context \cite{fairnessfails, fairnessabstraction}, the disconnect between being motivated by disparate impact and developing anti-racist solutions \cite{fairnesschoices}, and the disconnect between developer goals and algorithmic social impact \cite{Accountablealg}. 

The present paper urges collective reflection on research directions at this intersection. Can we do better to engage more intentionally and thoughtfully in this space? One problem when translating between (2) and (1) is that reconnecting back to either side is nontrivial. On the other hand, it is also nontrivial how to connect many algorithmic methods and streams of research to questions related to race and racism. For instance, scholars need more tools to connect questions in theoretical algorithmic fairness to applied machine learning research, let alone to design and sociology perspectives. Given that conducting research at intersections between subfields within computer science presents its own challenges, it is expected that intersections between entirely separate disciplines require particular tools. Though the intersection between algorithms and racism has proven to be meaningful in terms of sociopolitical relevance, it is indeed quite challenging in practice. 

In a recent \emph{Nature} paper, scholars leverage this distinction between overt and covert racism to reveal that language models can exhibit dialect prejudice against speakers of African American English (AAE) even when the model's overt stereotypes about African Americans are more positive on average~\cite{hofmann2024ai}. \citet{hofmann2024ai} also show that existing methods to alleviate racial bias in language models can amplify the discrepancy between covert and overt racism by obscuring racist attitudes maintained by the underlying model. This is an example of how understanding color-blind racism (i.e. the covert avoidance of racial language while maintaining racist outcomes) can lead to more nuanced research questions related to algorithmic bias. In this case, the language model's perpetuation of color-blind racism is shown to impact decisions made by that model. Hence, it is critical that researchers at the intersection of algorithms and racism establish a better understanding of color-blind racism to uncover important research directions that could remain unexplored otherwise.

In this paper, we connect the ideology of color-blindness to research in algorithmic fairness in an effort to strengthen the research agenda at the intersection between algorithms and racism. Here, color-blind racism is an ideology used to explain away racial inequalities under the guise of neutrality~\cite{10.1093/sf/soz162, bonilla2013racism}. \textbf{Despite being primarily motivated by racism, research in algorithmic fairness remains mostly disconnected from scholarship on racism. In particular, to the best of our knowledge, there has not been an examination connecting algorithmic fairness discussions directly to the ideology of color-blind racism; we aim to fill this gap.} 

We have organized the paper as follows. In \S~\ref{sec::colorblind-racism}, we begin by defining color-blind racism based on an account detailed by \citet{bonilla2013racism}\footnote{As noted in our limitations section, we have limited our discussion to this single account of color-blind racism. We justify that choice given the overall influence the account has had and continues to have in social science disciplines. Further, notable papers, e.g.~\cite{hofmann2024ai}, similarly use this reference as the primary source and motivation for connecting covert racism to algorithmic bias. Yet, they often assume the readers have a detailed understanding of color-blind racism, which is not often the case for researchers in algorithmic fairness; hence, part of our motivation for the present paper.}. These concepts highlight the complex nature of racial discourse and how ideological maneuvers function as roadblocks and detours while researchers navigate the intersection of color-blind racism and algorithms. In \S~\ref{sec::colorblindandalgorithms}, equipped with the account of color-blind racism, we distill a few guiding principles for research practices at the intersection. Further, we discuss concepts that make the scientific context complicated to contend with in practice. In \S~\ref{sec::fairness}, we dive deeper into the navigation experience by focusing on existing discourse within algorithmic fairness and the influence of color-blind racism. We discuss how understanding concepts from color-blind racism helps improve navigation by enabling the recognition of ideological shifts, disorientation, and disconnects. In \S~\ref{sec::limitations}, we detail various limitations of our work, including our limited scope, as well as opportunities for future work. Finally, in \S~\ref{sec::concluding} are concluding remarks. Ultimately, our goal is to spark discussion on the topic and advocate for maintaining deeper connections at this intersection.

At a high level, these concepts are related to previous work on the legal ramifications of algorithmic discrimination~\cite{DiscriminationinAi}. For instance, scholars have underlined problematic assumptions involved in treating any ``race-aware" algorithm as discriminatory and engaged anti-discrimination law to challenge these assumptions from a legal perspective~\cite{raceaware}. Our work is complementary to these ongoing discussions about the ways racism might be misunderstood or misinterpreted in the context of algorithms, although we do not contend with the legal view directly.

\section{Color-blind Racism}
\label{sec::colorblind-racism}

In this section, we discuss concepts that highlight the complex nature of modern racial discourse. As a starting place, we outline terms that may be useful in generating and orienting discussion at the intersection. Understanding patterns within racial ideology can help guide our navigation of the interdisciplinary landscape we occupy. A central challenge in this task of navigation is that modern racism is difficult to identify and address given that it is often concealed by seemingly ``race-neutral" methods and rhetoric. How do we begin to solve a problem that we cannot see?

Here, we consider color-blind racism as a powerful ideology that justifies racial inequities under the guise of neutrality~\cite{bonilla2013racism}. \citet{bonilla2013racism} outlines how color-blind racism manifests in several ways on an individual and societal level: 
\begin{enumerate}
    \item The tendency to avoid directly engaging with racial discourse.
    \item The avoidance of racial language while communicating about social problems that racial dynamics have historically shaped. 
    \item The invisibility of mechanisms that reproduce racially disparate outcomes.
    \item The oversimplification of the pursuit of racial justice to the search for outright ``racist'' actors who have an intent to cause harm.
\end{enumerate}  

We rely on the account put forth in \citet{bonilla2013racism} as an especially useful resource among other sources helpful in framing this problem\footnote{We do not claim to cover the full breadth of the ideology or the book {\em Racism Without Racists}. Our summary focuses primarily on chapters 1 and 3. We encourage readers to read the book in full. Also, the book demonstrates that the frames of color-blind racism are endorsed chiefly by white Americans and that black Americans are much less likely to use the frames directly. Even so, the book argues that color-blind racism is the dominant ideology shaping racial discourse in the US. And is, therefore, in our view, still worth considering widely.}. The representation of color-blind racism from \citet{bonilla2013racism} may add to our collective vocabulary as well as our understanding of racial discourse.

\subsection{Road Blocks} 
At the intersection of algorithms and racism, researchers face road blocks that lead to conceptual and ideological disconnects. For instance, scholars describe how socio-political and historical context tend to be disconnected from algorithmic fairness research questions~\cite{fairnessfails, fairnessabstraction} or how studies motivated by addressing disparate impact struggle to produce anti-racist solutions~\cite{fairnesschoices} or how the goals of technical developers can be disconnected from the social impact of the deployed models~\cite{Accountablealg}. Our aim is to better identify and anticipate these road blocks by examining the types of road blocks that arise in broader racial discourse.

In color-blind racism, people can leverage some concept in an effort to abstract away racialized context resulting in disconnects that are difficult to counteract. \emph{Abstract Liberalism} entails the abstraction of liberal ideals to explain racial matters, enabling one to appear logical and moral while actively opposing attempts to address racial inequality~\cite{bonilla2013racism}. For example, a common abstraction of ``equal opportunity'' tends to remove the socio-political context of who has been excluded from opportunities historically. Further, this abstraction diverts the conversation away from progress by shifting the question to whether historically disenfranchised and excluded groups are receiving ``preferential'' treatment, while ignoring patterns of under-representation, limited access to opportunities, etc. This diversion illustrates the power of abstract liberalism as a function of color-blind racism.

\subsection{Detours}
Researchers navigate intricate detours. Even when context remains connected, research questions can become sidetracked by maneuvers from color-blind racism. We need to be able to identify when and how this occurs in order to ensure we support proper grounding and re-connections. 

Consider the frames of color-blind racism that often function as detours in reasoning about racial inequality. \emph{Naturalization} involves positioning instances of racial inequality as natural occurrences~\cite{bonilla2013racism}. For instance, a naturalization view on racial residential segregation might question whether people naturally tend to group amongst others like themselves. \emph{Cultural racism} is a framing that positions persistent racial inequality as a result of cultural differences of racial groups~\cite{bonilla2013racism}. This positioning assumes racial inequality to be the result of the cultural inferiority of racial minorities. Cultural racism can be thought of as an evolved version of the biological inferiority frame well-known for older forms of racism. The extension moves away from the assertion that racial minorities are inherently inferior due to their biology and instead towards explaining inherent inferiority through the cultural differences between different racial groups. 

Lastly, \emph{Minimization of Racism} reduces the weight assigned to the historical context of racial inequality~\cite{bonilla2013racism}. Instead of completely ignoring or discounting the context of racial inequality, a slight concession is made. The idea is that racial discrimination is no longer a primary factor materially impacting people's lives but that it plays some less significant role. This minimization allows one to align with support for racial minorities by acknowledging the existence of racism while simultaneously arguing that race does not play a significant role in lived experience. Its power is in subtlety.

\subsection{Discussion} 

The frames of color-blind racism can make scholars hesitant and avoidant when navigating the intersection of race and algorithms. And even when engaging in racial discourse, we will see the influence of color-blind racism in making connections difficult to restore across disciplinary lines.

Unfortunately, these are powerful maneuvers that researchers need to be aware of and to which they should actively respond. They play a critical role in how we think about racial matters. For instance, it is difficult to respond when someone invokes the principles of liberalism to advocate against progress partially because the historical context of the problem has been removed. If one cannot recognize that this abstraction has occurred, it is much harder to reconnect to the original problem of racial inequality. We end up in cyclic streams of logic where we might find ourselves unsure how to respond to seemingly neutral attempts to justify various types of unfairness. And the maneuvers seem so reasonable and logical that one cannot trivially correct it. 

Fundamentally, these maneuvers can distort how we define and discuss racial discrimination. Racism becomes legitimate or relevant only in its ``all-out'' explicit form. Similar to procedural notions of colorblindness, we can become consumed by the outdated task of finding racist actors. The reasoning follows then that there must be evidence that every racial minority has been discriminated against in the same way at the same point in time for racism to be a relevant factor. Anything other than this is unreasonable. And if there exists a few outliers or exceptions (racial minorities) who do not experience the particular racial discrimination in question, then race must not be a significant factor by this logic. 

Notice that this understanding of racism resembles that seen in the Jim Crow era, where racist policies were applied unilaterally to all black people in the Jim Crow South. One problem with this view of racial discrimination is that the post-civil rights shift towards more institutionalized and systemic racism is more nuanced and complicated than the ``all-out'' racial discrimination of previous eras. Hence, once the logic of color-blind racism is invoked, our focus pivots to either searching for outright racist agents (demonstrated racist intent) or defending the value of addressing racism altogether. Neither directly supports the development of meaningful solutions to address systemic racism.

\section{Color-blind racism and Algorithms}
\label{sec::colorblindandalgorithms}

In this section, we take our understanding of the frames of color-blind racism~\cite{bonilla2013racism} and distill from this account a few guiding principles for ongoing research practices at the intersection between racism and algorithms. Here, we refer to ``scholars" as individuals conducting research at this intersection with particular emphasis on those approaching the intersection from the algorithmic perspective. Our aim is to outline considerations we find particularly relevant when color-blind racism encounters the scientific context. These considerations are aimed to support researchers who are conducting research at this intersection grappling with topics and concepts that have been shaped or influenced by racial inequality.

We underline the following principles grounded in ~\citet{bonilla2013racism}'s account. First, scholars should identify and challenge practices that disregard racialized context or racial issues. Second, scholars might consider dedicating more attention to lived experience and perspectives of minoritized people. Third, scholars should be attentive to harms related to a historical legacy of discrimination. To illustrate the value in each of these guides, consider each more closely:

\begin{enumerate}
    \item \textbf{Identify and challenge the disregard for racial context or explanations:} Scholars might unconsciously dismiss racial disparities or deny the existence of racism. This could take the form of inattention or unwillingness to reflect on racial issues compared to other axes of oppression. Scholars might internalize a view that race is not a significant societal factor or that racial progress has been achieved. This lack of concern tends to be a default response rather than a conscious cognitive belief.

    \item \textbf{Avoid the misevaluation of lived experience and perspectives of minoritized people:} Scholars fall somewhere on the spectrum between under-valuing and over-valuing lived experience. Undervaluing in the sense of dismissing perspectives of racial minorities as irrelevant and over-valuing when neglecting to account for disciplinary positioning. We need a balance. Storytelling and lived experience can play a role in research perspective. For instance, \citet{10.1145/3313831.3376392} highlights how storytelling and valuing the perspectives of racial minorities is critical for making the scientific community more inclusive. Leveraging lived experience and storytelling can serve as a powerful methodology in improving anti-racist efforts in our field. However, we also note that lived experience of individuals researchers should not be conflated with disciplinary expertise; When forming interdisciplinary collaborations, we should not naively assume all racial minorities are able to provide extensive perspective on race and racism scholarship. Many researchers racialized as minorities have received training only from the algorithmic perspective. In terms of the intersection between racism and algorithms, scholars need to consider this evaluation carefully.

    \item \textbf{Attend to de-prioritization of ongoing harms caused by a legacy of racial discrimination:} A more subtle form of ignoring context, scholars might acknowledge racial context but dismiss it as irrelevant to more recent occurrences. For instance, the belief that race is not a priority or less valuable to consider because of social progress is related to the minimization of racism. It can be a failure to recognize or engage with the existence of racial disparities and the need for ongoing efforts to address them. %
    
\end{enumerate}

Now, we extend this discussion of principled approaches to examine what makes this coupling of color-blind racism and algorithms complicated and deserving of careful attention. Said another way, we offer perspective on why it might be difficult to follow these guides in practice.

The concealed nature of color-blind racism complicates the role of intent. There can be disparate impact even without intention. Further, there can be racial inequality that arises even out of neutral or good intention. In essence, one can subscribe to a color-blind racist ideology without racist intent. This is one of the most important aspects of understanding color-blind racism. It requires scholar to have a more nuanced understanding of what it means to be neutral. In contrast, consider more simplistic notions of racism where one might say that any mention of race is racist. One reason for the confusion is that people often interpret the mere mention of race as suggestive of an intention. Similarly, admitting or advocating for being aware of race might be misinterpreted as suggestive of having discriminatory intentions~\cite{raceaware}. Color-blind racism offers an ideological framework that does not require racist intent; it is the real world impact that matters separate from intention.

Questions of objectivity arise naturally in the context of a scientific discipline~\cite{popper1972objective, kuhn1962structure}. The computer science perspective is thus influenced by this tendency to strive for more objectivity~\cite{philosophies6010022, sep-scientific-objectivity}. It is possible that race neutrality could be thought to align with scientific objectivity. Therefore, it can be confusing to think about color-blind racism as a scientific researcher. One might wonder: how can striving towards neutrality be problematic? We urge researchers against assuming that race neutrality is in alignment with scientific objectivity.\footnote{We briefly acknowledge possible relationship between scientific objectivity and discussions of race neutrality in the context of algorithms. A more detailed discussion is outside the scope of the present paper.} Recall how color-blind racism functions to render racial inequality neutral and reasonable. For this reason, color-blind racism might be more prominent amongst scientists and therefore more difficult to address.

Worth noting, race neutrality or color-blindness can range between passive and active. Active color-blindness involves actively blinding oneself to race. Passive color-blindness involves unconsciously disregarding or ignoring race. This range presents a challenge to the idea of color-blind racism: what do we make of instances where active color-blindness might in fact be the goal (procedural color-blindness~\cite{falsepromise, raceaware})? We revisit this discussion more in depth in the next section.

\section{Algorithmic Fairness}
\label{sec::fairness}

In this section, we focus on discourse within algorithmic fairness research and examine how this discourse has been influenced by color-blind racism. The four frames of color-blind racism typically function as ideological tools that influence the way we reason and discuss racism in the United States. It is not as simple as identifying each frame in isolation. In fact, the frames can be interwoven to form a protective boundary around racism. Notice the power in using liberal ideals to abstract away racial explanations and then pivoting to questions around racial privilege. The effectiveness of using minimization of racism as a starting point and therefore placing the burden on scholars to dedicate time to the task of defending the relevance of racism altogether. We will see these shifts in algorithmic fairness research discourse. For instance, the move from the question ``how do we measure the racial disparities of algorithms?" to the question ``are race considerations legal or fair in the first place?". How did we start at a conversation about correcting historical trends of racism and end up in a defensive position clarifying that correction is not equivalent to preferential treatment? The ideology of color-blind racism helps understand these pivots and disconnects. And further helps to anticipate shifts based on patterns in existing racial discourse. For instance, we might expect that the question ``how do we ensure a healthcare algorithm does not discriminate against black patients?" will shift to the question ``are race-aware healthcare algorithms unfair to white patients?".     

Consider the following analogy where we envision the landscape at the intersection between algorithms and racism like researchers as agents navigating various paths. If we think about a researcher as an agent navigating the disciplinary space between algorithms and racism, imagine dropping an agent onto the board and observing. Some of the agents begin squarely within the disciplinary lines of computer science but their destination is located across the other side in the land of race and racism. When we observe their trajectory, we notice a problem. They can orient themselves towards the other side. They can even begin walking in that direction, getting closer and closer. But for some reason, their plight gets disrupted. It is as if someone is intervening in the game. At times, it seems as though the agent is literally picked up and turned around by some outside force. At other times, we find the agent in some circular pattern far away from the destination. Or they are on a path of infinite internal spiral with not much connection to either side. It seems that they have lost sight of the original plan altogether. We can start some agents nearer to the middle of the intersection and observe how they navigate the space. Many times they end up helping to correct the paths of those around them and essentially building bridges and roads. They are sometimes trying to identify where the spirals and disconnects occur and urge the community to avoid those areas. And still other agents begin squarely with disciplinary training focused on race and racism. We might see these agents also struggle to find entry points of connection to the other side. 

In this imaginary view, there are some hidden forces at play. If the agent paths represent research discourse, then color-blind racism is acting as an invisible force influencing these paths. And the frames of color-blind racism function embedded in racial discourse continue to influence pivots, disconnects and diversions. This is partially because the agents operate without an awareness of color-blind racism to begin with. If there was a general understanding of the complicated nature of modern racial discourse then agents would have a lens through which to interpret and adjust ongoing discourse accordingly. 

 Equipped with an awareness of color-blind racism, we can improve the navigation of the landscape at the intersection between algorithms and racism by recognizing ideological shifts as such and re-orienting towards the practice of remaining meaningfully connected across disciplinary lines. Perhaps, we gain insight into the roles different researchers play in the discussion. Without being able to identify color-blind racism, these disconnects mimic the distortion and concealing of racism prevalent in racial discourse more broadly. Let us examine these disconnects more closely in the context of algorithmic fairness.

\subsection{Disconnection between Racialized Concepts and Oppression} \label{intersectionality}

In algorithmic fairness, racialized concepts often become disconnected from their original relationship to oppression and power. One example is the misinterpretation of the term ``intersectionality". Intersectionality is the unique combination of social and political identities resulting in distinct discrimination or privilege. The term has roots in black feminism and legal scholarship where black women needed to establish a unique form of discrimination not quite captured by previous precedents for black men or white women independently~\cite{Crenshaw, collective1977black}. When this concept entered the landscape of algorithmic fairness, the dominant interpretation partially reduced the term to simply mean combining identity categories; abstracting away the function of systems of oppression~\citep{kong2022intersectionally}. \citet{kong2022intersectionally} describes the dominant interpretation of intersectionality within algorithmic fairness as aiming for demographic parity across subgroups of race and gender. They present three weaknesses of this approach, one of which we will emphasize here: the concept of intersectionality addresses the compounding impact of overlapping systems of oppression, and the dominant interpretation in algorithmic fairness does not meaningfully address or challenge the unique nature of oppression that those at the intersections of identities face. Rather, the misinterpretation abstracts how systems of oppression function at all, and reduces this term to mean combining identity categories. While the term, intersectionality, has been misinterpreted in other fields, the setting of algorithmic fairness is particularly interesting. 

In the same way in which instances of racial inequality become detached from racism in the ideology of color-blind racism, racial concepts are being detached from meaningful origins in discourse within algorithmic fairness. There is a type of filtering effect. You take a concept like intersectionality, run it through racial discourse in algorithmic fairness and come away somehow having dropped the considerations of power and oppression. How does this happen? Bringing in the lens of color-blind racism is one way to pay closer attention to the filter itself.

Ultimately, the discussion shifts towards resolving the very problem the filter created. \citet{kong2022intersectionally} points out the research dilemma that arises post-filter between a focus on (i) splitting into smaller and smaller subgroups or (ii) prioritizing subgroups arbitrarily. The question ``does some unique combination of social and political identities result in distinct algorithmic discrimination?" turned into the question ``is it computationally feasible to consider all possible subgroup combinations?" or the question ``how should a particular combination of subgroups be prioritized?". What role is work like \citet{kong2022intersectionally} playing here? When the fundamental research questions have become disconnected from the conceptual framework they were born out of, the work is to highlight the disconnect and suggest paths towards re-connection. This is the role of this type of work. At such a complicated intersection, the importance of work that essentially re-orients and re-connects cannot be overstated. It is comparable to the role of signage and guides at physical intersections.

At times, this type of work might seem overly self-critical. We wonder why it seems scholars in algorithmic fairness tend towards internal critique where many papers take the tone of ``here is what we did wrong" or ``here is another misinterpretation" or ``here is something all these papers missed". Though it may seem like research in algorithmic fairness critiques simply for the sake of critiquing, there is something more important going on. Returning to the analogy where agents navigate a landscape, we see this work as a form of course correction and road building particularly at the intersection of race and algorithms. For every step we take towards algorithmic advances in this landscape, we need to re-evaluate the connection to racism. The aim is not to disrupt the progression of research but to enhance it via re-orientation and re-connection as research progresses.

Scholars navigating the overlapping landscape ought to anticipate an iterative process that could take the following form. Step 1: translate between racism and algorithms. Step 2: identify what has been gained and what has been filtered out. Step 3: return to step 1 to meaningfully re-establish the connection and repeat. We may find that the abundance of self critique is necessary precisely because of the intersection we navigate.

When translating a concept from racism to algorithms, connections get filtered out. When engaging in racial discourse, we encounter all the things that make racial discourse difficult in general. And color-blind racism is a helpful framework for understanding racial discourse from an ideological perspective offering many clues about how the disconnects occur and the types of ideological maneuvers to look out for.

\subsection{Implications of Fairness Metric Choice}
In algorithmic fairness research, scholars have developed metrics that capture notions of fairness from both the individual and group perspective for classification problems. This has led to debates around metric choice and ethical differences between different options. \citet{mismeasurefairness} underline important statistical limitations of fairness definitions while suggesting a useful categorization for types of fairness metrics. In one view, researchers require equal performance over protected groups via metrics that satisfy ``classification parity". In the other view, researchers examine how protected attributes impact decisions (related to the legal concept of disparate treatment~\cite{goel2017combating}). Interestingly, researchers have shown that it is impossible to satisfy certain fairness metrics simultaneously (demographic parity, equalized odds and calibration when base rates differ across subgroups)~\cite{doi:10.1089/big.2016.0047, impossibleKlenberg}. The limitations of different fairness metrics has garnered lots of discussion. For instance, \citet{kasy2021fairness} argue that some fairness definitions legitimize inequality by encoding a notion of merit in algorithmic design decisions; they suggest researchers refocus on the causal impact and challenge assumptions around choosing the objective function of the algorithm. 

One topic in algorithmic fairness discourse is whether researchers should frame fairness from the perspective of equity or equality. Equality typically involves treating all individuals the same in some sense without considering protected attributes. Whereas equity might include considerations of existing racial hierarchies and the possibility that different individuals might require different treatment to correct for historical inequalities. For instance, \citet{algreparations} see fairness metrics as flawed underlining their failure to address systemic issues and ultimately advocate for more ``reparative" algorithms based on an equity framing. 

Elements of abstract liberalism offer interesting insights into this conversation. Scholars often assume classification parity is a reasonable notion of equality partially in alignment with abstractions of equal opportunity. Countering the logic of color-blind racism would caveat that the existence of racial hierarchies in a given domain ought to inform the notion of equality defined. This is a point that is largely accepted within fairness discourse where researchers understand that deciding between fairness metrics must be guided by the downstream decision making process or social policy goals for a particular problem~\cite{mismeasurefairness}. Color-blind racism supports this idea and further clarifies that researchers should challenge assumptions around whether a notion of fairness might constitute neutrality or colorblindness. And ultimately whether that ought to be the goal. 

Scholars examining the legality of various fairness metrics point out that a useful dichotomy to consider is whether metrics are bias preserving or bias transforming~\cite{wachter2021bias}. The idea is that there may be a need for more transformational fairness metrics and methods that do not reinforce the status quo and reflect the biases prevalent in a social domain. Though it is not immediately clear what exactly bias transforming means from a technical perspective. And scholars have even argued that existing fairness metrics will not suffice in this shift towards fairness that incorporates equity and justice~\cite{mythofmethodology}. 

Underlying this discussion about fairness metrics is discourse around the ultimate goal of fairness research. Researchers grapple with how, when and why to specify fairness. Not all of this is directly tied to racism. But the discussions of equity and justice are largely motivated by instances of racism. And the scholars re-orienting and challenging the discussion are often leaning heavily on concepts from scholars in racism. Regardless of intentions or explicit rhetoric, many scholars are engaging in racial discourse. Hence, we see familiar themes of disconnect, disorientation and confusion arise; further underlining the importance of improving our collective understanding of modern racism.

\subsection{Building Models that Reflect the Racism of the World}

In one line of discussion at the intersection of race and algorithms, researchers consider that when algorithms are introduced in a particular social domain, the model output may simply reflect persistent inequality. For instance, in a standard classification task where training data is fundamentally flawed, model performance is prone to the same flaws. One of the most meaningful examples has been the use of re-arrest data as a proxy for recidivism~\cite{tschantz2022proxy}. Without considering the greater context of how racism connects to the criminal justice system, especially the system of policing in the United States, a model trained on re-arrest data is likely to reflect the over policing of black and brown communities. Essentially, begging the question: what are our goals when introducing algorithms into social domains? On the one hand, racial bias output by algorithms could be a useful indicator that racial bias persists in that domain. This is interesting in light of what we know about the concealed and hidden nature of color-blind racism. As racism becomes more complex and more difficult ``to see", might the role of algorithms be related to helping make visible that which has been hidden? There are policy considerations for what it means to preserve bias as an indicator in this way~\cite{wachter2021bias}. 

On the other hand, some scholars see an opportunity for algorithms to go further than humans have been able to. \citet{falsepromise} point out that the adoption of risk assessments has been motivated by two assumptions: i) the idea that assumed objectivity of algorithms could decrease human bias and (ii) the idea that algorithms could have potential to promote a new type of reform. Ultimately, they demonstrate the failures of these assumptions and challenge underlying ideas of color-blindness i.e. aiming to be race neutral in an unequal society will inevitably result in disparate outcomes~\cite{falsepromise}. Hence, scholars have pointed out the issues that occur when color-blind racism is ignored or when algorithmic fairness methods progress without deeper interdisciplinary insight~\cite{GreenandViljoen2020}. Would scholars more familiar with systemic racism have anticipated this happening? In fact, some did. \citet{Dirtydata} detail potential civil rights violations that occur from building algorithms on top of data that captures ``dirty policing".  And in terms of software development, \citet{SanchezRamos2023} conduct a mixed-methods study on how personal biases impact algorithms in relation to color-blind ideology expressed by computer science undergrads at Hispanic-serving institutions. In these examples, the research supports connecting algorithmic fairness discussions to concepts from color-blind racism in an effort to clarify and or re-direct research discourse. 

We see that this discussion has been enhanced by the studies dedicated to revealing hidden assumptions and re-interpreting technical results. Reiterating the importance of actively forging interdisciplinary connections between research at the intersection of algorithms and racism. And we can see how discourse about the goals and promises of algorithmic fairness resemble some of the trends in racial discourse more broadly.

\subsection{Race-Neutral Rhetoric} 
Within the field of algorithmic fairness, there has been a shift towards race-neutral language despite work being largely motivated by racism. For instance, scholars often generalize by moving from the word ``racism" to ``bias"~\footnote{To be sure, there are streams of research in algorithmic fairness where this move is warranted. Our goal is simply to urge awareness around rhetorical shifts and encourage researchers to be more specific when possible.}. Consistent with the guiding principles for ongoing research practices, we advocate for more specific language when studying at the intersection of race and algorithms. The tendency towards race-neutral language in the field of algorithmic fairness, particularly when engaging in racial discourse, is partially due to the fact that notions of fairness are applicable to various axes of identity. And of course, algorithmic fairness is not only focused on algorithms and racism. In instances where the research focus is more broad, the term ``bias" is often appropriate. We are more interested in scenarios where the only type of bias articulated as motivating a study is racism. Where the shift to a more general term has advanced a socio-technical disconnect that will need correcting in the future. Where the appeal of a more general term is naively to be race-neutral. Plainly, scholars in algorithmic fairness discussing racism ought to state this more directly resisting the urge to avoid racial rhetoric. Per our discussion of color-blind racism, concealing racism ultimately makes it more difficult to contend with. Instead of shifting to race-neutral language, we advocate for being explicit about the motivation of the work and the intended impact of the work moving forward.

\section{Limitations and Future Work}
\label{sec::limitations}
We have discussed racism primarily from the perspective of research motivated by addressing anti-black racism in the United States. In doing so, we have not incorporated wider ethnic studies perspectives focused on racism as it relates to non-Black racial minorities. We acknowledge this as a limitation, however we also justify this choice. Grounding in anti-black racism in discourse around racism can lead to advancements not only for black people but for other minority groups i.e. the benefits of desegregation~\cite{brooks2009alien}, affirmative action~\cite{cortese1991affirmative} and other advancements in the U.S. designed to benefit black people that ultimately benefit many others.

We have discussed racism as an ideology, which has its limitations. \citet{garcia_notes} outlines several considerations with accounts of racism as an ideology. While racism as an ideology can be helpful, it is important to examine non-ideological accounts in future work. Further, there are ongoing philosophical, sociological and legal debates over the ongoing causes of racial disparities that should be considered. For instance, \citet{darby2010} argues for stronger normative arguments that stand regardless of how racial inequalities are explained. In future work, these types of perspectives can strengthen the discussion of discourse at the intersection of race and algorithms.

The theory underlying the account of color-blind racism builds on work that is mostly based on American racism. This focus on the US context means this paper does not fully contend with racism outside the US that can have similarities but deserves independent consideration. We have also limited our discussion to racism as the axis of oppression; other types of harms and biases are outside the scope of this work. And lastly, we understand that race is socially constructed and still has a meaningful impact on our lives throughout society. As mentioned, part of our motivation is to ensure our understanding of racism is moving forward as technology influences and transforms racial inequality. 

We have focused on one account of color-blind racism~\cite{bonilla2013racism} within the broader context of discourse on the topic. For instance, Bonilla-Silva has a large body of work on the topic that we do not discuss in depth~\cite{bonilla2013new, bonilla2011sweet, bonilla2022color, bonilla2015structure} (non-exhaustive). In future work, building on this initial discussion, it will be important to expand the discussion of the intersection between color-blind racism and algorithms beyond this particular account and incorporate additional perspectives. In the present paper, we limit our scope. Let us state directly that our contribution is not intended to be a comprehensive review across all research on color-blind racism. We view this as a growing area that is ripe for further exploration.

This work is related to the conceptualization of racism in scientific research. For instance, \citet{lett2022conceptualizing} outline the failure modes associated with the intersection between race and quantitative health sciences research. Also, \citet{valuesML} detail a value system within machine learning research showing that the dominant values in applied machine learning disproportionately neglect historically marginalized people. Our work is complementary to this stream of research. But much more work is needed to fully examine the dynamics between researchers at the intersection of race and algorithms.

We have discussed racism while mostly assuming unintentional or unconscious racism. Yet the possibility of racism existing without racists~\footnote{The title of \citet{bonilla2013racism} is ``Racism Without Racists: Color-blind Racism and the Persistence of Racial Inequality in America''} explains that institutional structures tinged with their creators’ or designers’ racism can persist and perpetuate racial disparities even after the originators have passed away. In essence, there can be no racism at any time without there having been racists earlier, and the harm caused by institutional, structural, and systemic racism can be widespread and significant. \citet{bonilla2013racism}’s influential trope of ``racism without racists'' helps remind us that when racism is translated into institutions, it must be translated from something. Future discussions might contend more deeply with the role of intent.

\section{Concluding Remarks}
\label{sec::concluding}
To strengthen research at the intersection between algorithms and racism, we have reviewed the frames of color-blind racism and connected them to research in algorithmic fairness. Recall that color-blind racism is an ideology used to explain away racial inequalities under the guise of neutrality. Functionally, color-blind racism shows up as an unwillingness to engage in racial discourse, avoidance of racial language while communicating about social problems shaped by racial dynamics, oversimplifying racial justice to a search for outright ``racists," and more. In the context of algorithms, a better understanding of color-blind racism can inform collective research practices. Namely, scholars can identify and challenge practices that ignore racial context or explanations, dedicate more attention to the lived experience and perspectives of minoritized people, and attend to harms related to a historical legacy of discrimination. 

We discussed points of confusion that arise, particularly when race neutrality might be conflated with scientific objectivity. And the vital distinction between passive and active notions of color-blind racism that may arise in the context of algorithms. Finally, we offer perspective on algorithmic fairness discussions through the lens of color-blind racism. We see color-blind racism as a valuable tool in understanding how racialized concepts become disconnected and misinterpreted in the context of algorithmic fairness, how color-blind racism surfaces in the discourse around fairness metrics and the overall goals of algorithmic fairness research, and how the shift toward race-neutral language deserves thought and attention as research progresses.

Overall, we advocate for drawing deeper connections between algorithms and racism. In addition to sparking discussion, this is a call to action for more thoughtful engagement at our intersection. This is not a call for abandoning disciplinary lines altogether; there is immense value in continuing to develop research on both algorithms and racism separately. This paper arises from the recognition that the intersection between algorithms and racism deserves more attention and more creative engagement. We should clearly state our attempts to do this translational work between fields, including the challenges and the limitations. In future work, we will need unique curriculum development dedicated to this intersection; the task is not as simple as training scholars in computer science or African American studies separately. There is a growing need for training people directly at the intersection that may require new courses, internships, job titles, etc. There may be untapped institutional potential associated with dedicating time and resources to examine algorithms and racism.

In conclusion, understanding color-blind racism is crucial for addressing the racial disparities that have motivated much of the algorithmic fairness research agenda. Recognizing its invisibility is essential in developing effective solutions to combat racial bias in algorithms. Ultimately, these simple guiding principles can help identify occurrences of color-blind racism to promote racial regard by attending to racism more directly at the intersection of algorithms and racism.

\clearpage
\section{Research Ethics and Social Impact Statement}

This paper does not conduct research analysis with users or rely on sensitive data. However, we did take care to consult experts and resources within algorithmic fairness, African American studies, moral philosophy and philosophy of science during the writing of this paper. We did this to ensure that our analysis is well-informed and properly scoped within the context of the perspectives discussed. Notice that the scope of the paper is quite narrow in focusing on a single account of color-blind racism, which is intentionally aligned with expert guidance.

Discussions of racism, particularly anti-Black racism, can provoke intense debates and feelings of discomfort. We anticipate that researchers who focus on algorithmic fairness solely from the technical perspective might express defensiveness or other strong emotions in response to the argument presented in this paper. Further, in public racial discourse, discussions of color-blindness or race neutrality are often weaponized against progressive political agendas. This potential misinterpretation underlines the importance of the limited scope of the paper (focusing on one account) and the diligence required in the development of this paper. 

Another important misinterpretation of this work could be the idea that if one applies the guiding principles suggested and also enhances their understanding of color-blind racism, then the work to contend with racism is done. This paper is not intended to support this conclusion. We see this account of color-blind racism as non-exhaustive. We see this paper as sparking important discussions, and as enhancing ongoing research development at the intersection. We do not see this single paper as the end all solution to contemplating racism in the field of algorithmic fairness. There is much work to be done and this will continue to be an ongoing process.

\section*{Acknowledgements}
Thank you to colleagues at Harvard for feedback and helpful discussions. Thank you to Camille Harris and Alexander Tolbert for editing support on previous versions of the paper. Thank you to Jennifer Chien, A. Feder Cooper, Hailey James, Shannon Whittaker and Harvard EconCS friends for feedback and helpful discussions. JWD is supported by a Ford Foundation Pre-doctoral Fellowship and the NSF  Graduate Research Fellowship Program under Grant No. DGE1745303. Any opinions, findings, and conclusions or recommendations expressed in this material are those of the author(s) and do not necessarily reflect the views of the NSF.


\clearpage 
\bibliographystyle{plainnat}
\bibliography{sources}


\end{document}